\documentclass[aps,prl,amsmath,amssymb,twocolumn,showpacs,showkeys,superscriptaddress]{revtex4-1}

\usepackage{graphicx}
\usepackage{dcolumn}
\usepackage{color}
\usepackage{ulem}

\begin{document}
\title{Type-II InAs/GaAsSb/GaAs quantum dots as artificial quantum dot molecules}

\author{P. Klenovsk\'y}
\email[]{klenovsky@physics.muni.cz}
\affiliation{Central European Institute of Technology, Masaryk University, Kamenice 753/5, 62500~Brno, Czech~Republic}
\author{V. K\v{r}\'apek}
\affiliation{Central European Institute of Technology, Brno University of Technology, Technick\'a 10, 61600 Brno, Czech Republic}
\author{J. Huml\'i\v{c}ek}
\affiliation{Central European Institute of Technology, Masaryk University, Kamenice 753/5, 62500~Brno, Czech~Republic}

\date{\today}

\begin{abstract}

We have studied theoretically the type-II GaAsSb capped InAs quantum dots for two structures differing in the composition of the capping layer, being either (i) constant or (ii) with Sb accumulation above the apex of the dot. We have found that the hole states are segmented and resemble the states in the quantum dot molecules. The two-hole states form singlet and triplet with the splitting energy of 4 $\mu$eV / 325 $\mu$eV for the case (i) / (ii). We have also tested the possibility to tune the splitting by vertically applied magnetic field. As the predicted tunability range was limited, we propose an approach for its enhancement.

\end{abstract}


\maketitle

\section{Introduction}
InAs quantum dots (QDs) on GaAs substrate covered by the GaAs$_{1-y}$Sb$_y$ capping layer (CL) exhibit a lot of interesting properties~\cite{Liu,LiuSteer}. The prominent one is the possibility to tune the type of the band alignment continuously from type-I to type-II by changing the Sb content in the layer~\cite{Kle,KleJOPCS,KleNOCON}. Furthermore, in type-II regime these dots exhibit a large blue-shift of the emission with increasing pumping~\cite{LiuSteer,SiGeKlenovsky,Kle}.\\
We have previously found that the hole wavefunction in the type-II regime resembles that of the quantum dot molecule (QDM) with the symmetric and antisymmetric state separated by the energy of $\sim 0.6$~meV~\cite{Kle}. QDMs might be important for the quantum information processing in the future, e.g.~as quantum gates (QGs)~\cite{Burkard}. Vertical QDMs can be easily manufactured and offer a strong coupling~\cite{Payette} and a tunable tunneling~\cite{QDmoleculesStrainTuningVK}, but they cannot be scaled and suffer from the internal symmetry lowering induced by the strain field~\cite{Bester}. On the other hand, existing lateral QDMs exhibit a weak coupling~\cite{Beirne} and their properties cannot be tuned easily because of the complex growth procedures~\cite{Wang}.\\
The properties of the QDMs under consideration would be in many aspects superior to the existing ones. Particularly, the distance between the segments of the hole wavefunction and the tunneling energy between them could be tuned by the size of the QD, the thickness of the CL and its Sb content in a certain range until the QDM collapses into a single elongated QD or separates into two independent QDs~\cite{KrapekNottingham}. Although only holes form the molecular states which rules out the applications requiring both electrons and holes, there exists a theoretical proposal for a quantum gate based on only one type of charge carrier~\cite{Burkard}. It requires the molecular states of two charge carriers with a singlet-triplet splitting tunable to zero.\\
\section{Methods}
We have calculated two-hole states first by obtaining the single-particle wavefunctions by the envelope function approximation using Nextnano simulation suite~\cite{next} followed by multi-particle states calculated by the method of configuration interaction (CI)~\cite{Stier1999}. We have varied the Sb content in the layer and the external magnetic field acting on the structure. The latter is needed in order to bring both singlet and triplet states to the same energy as required by the proposal of the quantum gate~\cite{Burkard}.
\begin{figure}[!ht]
\renewcommand{\tabcolsep}{2pt}
\begin{center}
\begin{tabular}{c}
\includegraphics[width=0.3\textwidth]{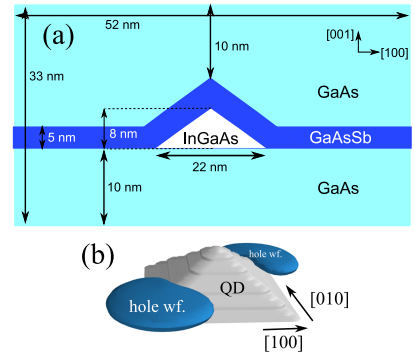} \\
\end{tabular}
\end{center}
\caption{
(a) Side view of the structure (i). At the bottom there was a 10~nm thick layer consisting of pure GaAs. On top of it the $\mathrm{In_xGa_{1-x}As}$ QD with trumpet-like In composition profile~\cite{Migliorato} was defined as a 8~nm high pyramid with its sides oriented along $[100]$ and $[010]$ directions. Above the QD the 5~nm thick $\mathrm{GaAs_{1-y}Sb_y}$ CL with constant $y$ was defined. The structure was finished by pure GaAs. Panel (b) shows the calculated probability density of the hole single-particle ground state. Note the two segments of the wavefunction next to the QD and the orientation of those along [110] crystal direction.
\label{figSbStruct}}
\end{figure}
\section{Results}
\begin{figure}[!ht]
\begin{center}
\includegraphics[width=0.3\textwidth]{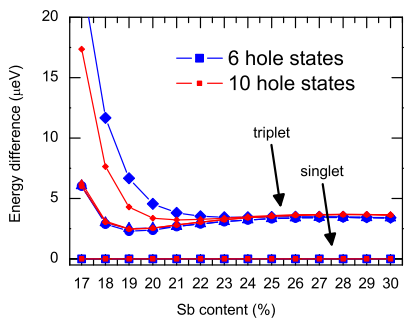}
\end{center}
\caption{Calculated $y$-dependence of the energies of the complex of two holes with respect to the lowest state energy. Either six (blue curve) or ten (red curve) single particle basis states were used. The calculated states are marked from the lowest to highest energy difference from the state with the lowest energy by squares, circles, triangles and diamonds, respectively. It can be seen that regardless of the number of basis states, for $y$ larger that $\sim 0.23$ singlet and triplet is formed, the former being lower in energy by $\sim 4\mathrm{\mu eV}$. These calculations were performed for the structure shown in panel (a) of Fig.~\ref{figSbStruct}.
\label{figSb12}}
\end{figure}

\begin{figure}[!ht]
\begin{center}
\includegraphics[width=0.3\textwidth]{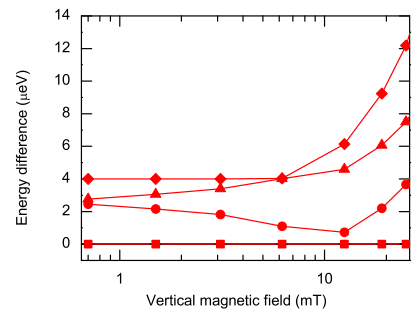}
\end{center}
\caption{Magnetic flux density ($B$) dependence of the energies of two-hole complexes (relative to the lowest one). Sb content in the CL was $y=0.25$ in this calculation and 10 hole single particle states were used as the basis for the CI calculation. The simulated structure is shown in panel (a) of Fig.~\ref{figSbStruct}. The magnetic field was considered only along the vertical axis of the QD. Only values of the energy difference for $B$ up to 25 mT are shown as for higher values of $B$ the triplet is completely dissolved. 
\label{figSb13}}
\end{figure}

The calculations were done for two structures with different distributions of Sb in the capping layer, i.e. (i) InGaAs QD capped by 5~nm thick GaAsSb CL with constant Sb content, see Fig.~\ref{figSbStruct} (a), and (ii) the same dot capped by the $\mathrm{GaAs_{1-y}Sb_y}$ CL with trumpet Sb content $y$~\cite{Migliorato}, ranging from 0.05 close to the dot base to 0.22 above its apex. Furthermore the CL was not replicating the dot, but was designed as a flat layer. Its thickness was 14~nm in the areas where the dot was not present and 6~nm above the QD apex. The particular choice of both structures is based on experimental studies~\cite{UlloaHomogSRL,Ulloa,Hospodkova2013,Zikova2015}.\\
First we discuss the structure (i) with the homogeneous Sb distribution in CL. The calculated single particle hole wavefunctions consisted of two segments oriented along [110] crystal direction~\cite{Kle}, see Fig.~\ref{figSbStruct} (b). 
We have found that the two-hole states in this structure form a singlet and a triplet by calculating the energy spectrum of the QDs containing two holes by the CI method~\cite{Stier1999} with basis states obtained by the envelope function approach~\cite{next}. 
In Fig.~\ref{figSb12} we show the energies of the four lowest states of two holes as a function of Sb content $y$. As $y$ increases, the energies of the three lowest excited states get closer together and finally at $y\approx 0.23$ their energy difference is less than 0.1~$\mathrm{\mu eV}$ which is within the error of the calculations. Thus, they form a triplet state. The energy difference to the singlet state is $\sim 4\mathrm{\mu eV}$.\\
The singlet-triplet structure can be represented with the effective Heisenberg Hamiltonian $H_s=J\vec{S_1}\vec{S_2}$, where $\vec{S_1}$ and $\vec{S_2}$ are the spins of the two holes present in the dot, and $J=\epsilon_t-\epsilon_s$ is the exchange energy. The variables $\epsilon_t$ and $\epsilon_s$ denote the energies of the triplet and that of the singlet, respectively. The exchange energy for structure (i) is thus $J=4\mathrm{\mu eV}$, i.e.~by more than two orders of magnitude smaller than that calculated by Burkard et.~al.~for two electrons in their double QD~\cite{Burkard}. Note that this value is already at the limit of the experimental resolution and leaves little space for its further reduction to zero as required by Burkard's proposal.\\
However, we have tried to verify whether $J$ can be tuned by applying magnetic field along the QD vertical axis as proposed in Ref.~\cite{Burkard}. The result can be seen in Fig.~\ref{figSb13}. By applying magnetic field we observe a steady increase of $J$.
Note that for higher B the Zeeman coupling of the spins to the magnetic field exceeds the exchange energy and the triplet state is broken. It would require a stronger $J$ to preserve the singlet-triplet structure in high fields.\\
It has been reported that the GaAsSb layer can be inhomogeneous with the antimony-rich region just above the top of the InAs QD~\cite{Ulloa2} which inevitably forces the holes to be located there. This is probably due to inhomogeneous strain distribution around the QD~\cite{KleJOPCS}. For the same reason, we expect that CL can be considerably thinner above the dot~\cite{Ulloa2}. We will now explore the consequences of the inhomogeneous antimony distribution and CL thickness for the two-hole states.\\
This scenario is modeled by the structure (ii), i.e. InGaAs QDs capped by GaAsSb CL with nonuniform Sb content. The calculated hole wavefunctions consist again of two segments. Now, however, the segments are oriented along [1-10] crystal direction and the whole wavefunction resides above the QD. The energies of the four lowest states of two holes in this structure, calculated by CI with 10 basis single particle states, formed a singlet and a triplet separated by the energy of $325\,\mathrm{\mu eV}$. Here it was not possible to reduce $J$ by magnetic flux density $B$ to zero, only a steady increase of $J$ with $B$ was observed.
\begin{table}[!ht]
\begin{tabular}{|l|c|c|c|}
\hline
& structure (i)& structure (ii)&Ref.~\cite{Burkard}\\
\hline
$J$ at $B=0$~T ($\mu$eV)& 4& 325& 700\\ 
$m$ ($m_e$)& 0.1& 0.1& 0.067\\
$a_B$ (nm)& 9.5& 6.3& 20\\ 
$a/a_B$& 1.6& 0.6& 0.7\\
$\hbar\omega_0$ (meV)& 8.5& 19& 3\\
$\tau$ (\%)& 0.5& 33.5& $\lesssim$20\\
\hline
\end{tabular}
\caption{Comparison of selected parameters between the structures (i), (ii), and calculations of Burkard et.~al.~\cite{Burkard}. The meaning of the parameters is the following: $m$ stands for effective mass; $a_B=\sqrt{\hbar/m\omega_0}$ is the effective Bohr radius; $\hbar\omega_0$ is the energy difference between the single particle hole state belonging to Bloch wave with s-symmetry and that with p-symmetry; $\tau$ is the ratio of the probability density right in the middle between the segments to the peak probability density.\label{tab_Comparison}}
\end{table}
The reason for insufficient tunability of both structures (i) and (ii) can be found by an inspection of the single particle hole wavefunctions and comparing the results with those of Ref.~\cite{Burkard}, see Tab.~\ref{tab_Comparison}. Particularly, the distance $a$ between the segments of the hole wavefunction relative to the effective Bohr radius $a_B=\sqrt{\hbar/m\omega_0}$ is $a/a_B=1.6$ for structure (i) resulting in a weak tunneling between the segments. Thus, the two segments in (i) act as two non-interacting QDs. The exchange interaction is too weak  and the singlet-triplet structure is broken before the field is high enough to induce sizeable changes. The structure (ii) is the opposite case. While the segments are closer together and $a/a_B=0.6$, their overlap is too large. The probability density right in the middle between the segments, denoted as $\tau$, is $33$~\% of the peak probability density. The structure therefore behaves like a single elongated dot rather than the molecule; s and p-like single hole states are formed instead of symmetric and antisymmetric molecular states~\cite{KrapekNottingham}. Note that the scenario proposed by Burkard falls somewhere in the middle between those of our structures. Thus, by optimizing the properties of our QD and CL a structure for which $J$ can be tuned to zero by $B$ might be designed. The tuning ``knobs" can be e.g. the QD size and Sb distribution in the CL. We hope that a progress might be achieved with smaller QDs than those considered in this work capped by thin CL with large uniform Sb content (e.g. up to $y$=0.3).\\
\section{Conclusions}
To conclude we have calculated the singlet-triplet splitting of two-hole states for type-II GaAsSb capped InAs QDs for two structures differing in the composition profile of the capping layer. The values of the splitting were found to be 4 and 325 $\mu$eV for the structures (i) and (ii), respectively. However, for neither Sb profile a possibility to tune $J$ to zero was predicted. This was explained by large / small separation of the segments of the hole wavefunction and resulting too weak / too strong coupling between them for the case of the structure (i) / (ii). Finally, strategies to overcome this shortcoming were proposed.
\section{Acknowledgements}
The work was supported by the project no. TH01010419 of the Technological agency of the Czech republic, the European Regional Development Fund, project
No. CZ.1.05/1.1.00/02.0068, and the European Social Fund, grant No. CZ.1.07/2.3.00/30.0005.

\end{document}